%% file: main.tex
\documentclass[9pt,twocolumn,twoside]{pnas-new}

\usepackage{graphicx}			
\usepackage{subfigure}
\usepackage{float}
\usepackage{amsmath,amssymb,bm}
\usepackage{mathtools}

\templatetype{pnasresearcharticle} 

\addto\captionsenglish{}

\title{Multiplicação de matrizes: uma comparação entre as abordagens sequencial (CPU) e paralela (GPU)}

\author[a,1]{Andre G. C. Pacheco}

\affil[a]{Programa de Pós-Graduação em Informática (PPGI), Universidade Federal do Espírito Santo (UFES), Vitória - ES, Brasil}

\dates{Este relatório técnico foi produzido em Novembro de 2016}

\doi{\url{Relatório Técnico}}

\leadauthor{André G. C. Pacheco} 

\correspondingauthor{\textsuperscript{1}Autor correspondente. E-mail: agcpacheco@inf.ufes.br}


\begin{abstract}

\input{src/abstract.tex}

\end{abstract}

\begin{document}

\maketitle
\thispagestyle{firststyle}
\ifthenelse{\boolean{shortarticle}}{\ifthenelse{\boolean{singlecolumn}}{\abscontentformatted}{\abscontent}}{}

\section{Introdução}
\input{src/intro.tex}

\section{Arquitetura de GPUs - uma visão geral}
\input{src/arquitetura-gpus.tex}

\section{Multiplicação de matrizes}
\input{src/mult-mat.tex}

\section{Experimentos} 
\input{src/experimentos.tex}

\section{Conclusão}
\input{src/conclusion.tex}

\section*{Referências}
\bibliography{references}

\end{document}

%% file: src/abstract.tex
A modelagem de problemas utilizando matrizes é de extrema importância para Ciência da Computação. Áreas como computação gráfica, grafos e aprendizado de máquina utilizam matrizes com alta frequência para solucionar seus respectivos problemas. Dessa forma, operar matrizes de maneira eficiente é muito importante para o desempenho de algoritmos. Uma das operações de matrizes mais utilizadas é a multiplicação, que se torna um empecilho para o desempenho computacional de algoritmos na medida que o tamanho das matrizes a serem multiplicadas aumentam. Por conta disso, a computação paralela se tornou uma solução padrão para abordar tal problema. Neste trabalho é apresentado uma comparação entre as abordagens sequencial e paralela para multiplicação de matrizes utilizando CUDA e OpenMP. O resultado da análise realizada entre o tamanho da matriz e o desempenho da multiplicação mostra a importância da paralelização principalmente para matrizes de ordem elevada.

%% file: src/intro.tex
Operações com matrizes são fundamentais para computação científica de maneira geral. Diversos algoritmos das mais variadas áreas da computação são modelados e executados utilizando o conceitos e operações matriciais \cite{kakaradov2004}. Uma operação matricial fundamental é a multiplicação. Tal operação é bastante comum em algoritmos de diversas áreas e possui um grande potencial de consumo de tempo computacional. Com a crescente demanda de dados, áreas como a de aprendizado de máquina (\textit{machine learning}), exigem algoritmos cada vez mais eficientes para que os mesmos sejam capazes de processar mais informações em menos tempo. Com isso, otimizar operações matriciais como a multiplicação é de suma importância para que esse objetivo seja alcançado. Uma abordagem padrão para otimização de processamento de dados é o uso de paralelismo computacional e uma forma eficiente e massivamente utilizada de paralelização é o uso de uma Unidade de Processamento Gráfico (\textit{Graphics Processing Unit} - GPU). Atualmente paralelização e GPUs se tornaram quase que sinônimos e diversos algoritmos, das mais diversas áreas da computação, utilizam GPUs para otimizar seu processamento \cite{sanders2010}.

A computação acelerada por placas de vídeo é o uso de uma GPU juntamente com uma CPU (\textit{Central Processing Unit}) para acelerar algoritmos de aprendizado, análise e engenharia de maneira geral. O uso de GPUs como propósito geral de computação foi introduzido pela NVIDIA corporation em 2006 por meio da plataforma de programação CUDA (\textit{Compute Unified Device Architecture}), que permite desenvolvedores codificá-la utilizando a linguagem de alto nível C \cite{manualCUDA}. Desde seu lançamento, a plataforma vem potencializando \textit{data centers}, universidade, empresas de médio e grande porte ao redor do mundo \cite{cook2012}.

Atualmente GPUs desempenham um papel fundamental na aceleração de aplicativos em plataformas que variam de inteligência artificial até carros, drones e robôs \cite{CUDAzone}. Ao longo dos anos, diversos trabalhos vêm sendo desenvolvido utilizando a plataforma CUDA. Jang et al. \cite{jang2008} apresentaram uma implementação de uma rede neural utilizando CUDA; Krizhevsky, Sutskever e Hinton \cite{krizhevsky2012} classificaram 1.3 milhões de imagens utilizando redes neurais profundas e computação paralela; Veronese e Krohling \cite{veronese2010} utilizaram a computação paralela aplicada à otimização por meio de algoritmo evolutivo. 

Uma outra maneira simples de se obter paralelismo utilizando apenas os núcleos de uma CPU é utilizando OpenMP, uma interface de programação paralela de memória compartilhada para arquitetura de múltiplos processadores \cite{chapman2008}. Sendo assim, neste trabalho a computação paralela utilizando GPU e OpenMP é realizada para computar a multiplicação de matrizes e seus resultados são comparados com versão sequencial utilizando a CPU. O restante deste artigo esta organizado da seguinte forma: na seção 2 são apresentados conceitos básicos relacionados a arquitetura de uma GPU; na seção 3 a multiplicação de matrizes é discutida; na seção 4 os resultados experimentais são analisados; por fim, na seção 5 é realizada uma breve conclusão.

%% file: src/arquitetura-gpus.tex
Nesta seção, são apresentados os conceitos básicos relacionados a arquitetura de GPUs e a plataforma CUDA.

\subsection{Arquitetura}
\begin{figure}[h]
\centering
\includegraphics[scale=0.4]{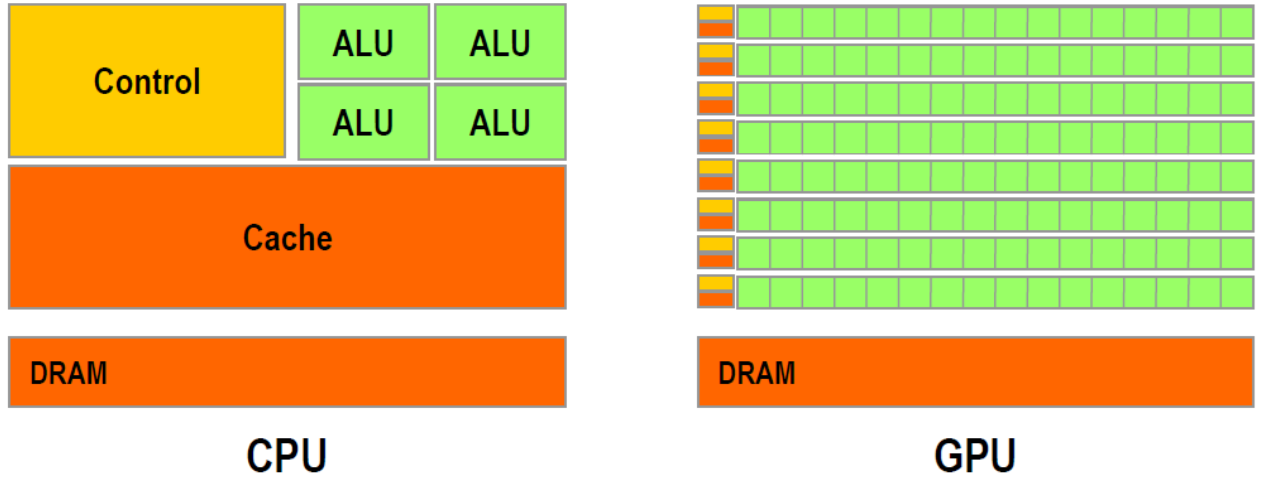}
\caption{Comparação entre CPU e GPU}
\label{fig:cpuvsgpu}
\end{figure}

GPUs são unidades de processamento especializadas em processamento gráfico da classe SIMT (\textit{Single Instruction Multiple Threads}), neste modelo, múltiplas \textit{threads} independentes são executadas de maneira concorrente utilizando uma instrução única \cite{manualCUDA}. Desde 2006, com lançamento da plataforma CUDA, GPUs vêm sendo utilizadas com propósito geral de computação. Uma maneira simples de compreender a diferença entre uma GPU e uma CPU é comparar o modo que as mesmas processam suas tarefas. Uma CPU possui alguns núcleos otimizados para o processamento serial sequencial, enquanto uma GPU tem uma arquitetura paralela gigantesca que consiste em milhares de núcleos menores e eficientes criados para lidar com múltiplas tarefas simultaneamente \cite{CUDAzone}. Como mostra a Figura \ref{fig:cpuvsgpu}, a CPU dedica grande parte dos seus circuitos ao controle, enquanto a GPU foca mais nas ALUs (\textit{Arithmetic Logic Units}), fazendo com que a mesma seja mais adequada para cálculos paralelos.

\begin{figure}[h]
\centering
\includegraphics[scale=0.4]{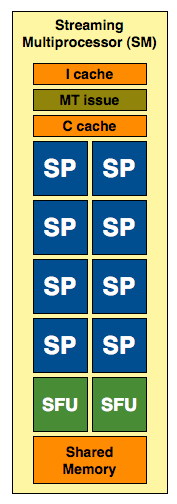}
\caption{\textit{Streaming Multiprocessor} de uma NVIDIA G80/G90}
\label{fig:SMs}
\end{figure}

Um pré-requisito para codificar em CUDA é conhecer a arquitetura das GPUs. Quanto mais conhecimento o programador adquirir, mais otimizados serão seus códigos. Existem diferentes arquiteturas de GPU, todavia existem diversas características comuns a todas elas. De maneira geral, GPUs da NVIDIA são divididas em SMs (\textit{Stream Multiprocessors}), local onde um conjunto de \textit{threads}, denominado \textit{warp}, é executado. Cada SM possui vários núcleos de processamento chamados de CUDA \textit{cores} (ou \textit{Streaming Processor - SP}), que por sua vez possuem \textit{pipelines} completos de operações aritméticas (ALU) e pontos flutuantes (\textit{Floating Point Unit} - FPU). Cada SM possui uma memória dedicada chamada de memória compartilhada (\textit{shared memory}). Essa memória é fisicamente próxima dos \textit{cores} e possuem acesso muito rápido, porém são pequenas (na ordem de KB). Além disso uma SM possui um \textit{cache} de instruções, um cache de constantes, uma unidade de funções especiais (SFU - \textit{Special Function Units}) e um bloco para escalonar \textit{warps} (\textit{Multithread Instruction Fetch and Issue Unit} - MT Issue). O número de SMs, \textit{cores}, tamanho de cache, dentre outros, podem variar de acordo com o modelo da GPU. A Figura \ref{fig:SMs} ilustra um SM de uma NVIDIA G80/G90, na qual possui 8 CUDA \textit{cores}.

\textit{Threads} podem ser agrupadas em blocos que executam na mesma SM compartilhando a mesma \textit{shared memory}. Cada SM possui um limite máximo de blocos, que por sua vez possui um limite máximo de \textit{threads}. Um conjunto de blocos criados por um código CUDA é chamado de \textit{grid}. Os valores dos limites também são parâmetros de cada modelo de GPU. 

As principais memórias de uma GPU são descritas a seguir:

\begin{itemize}
\item \textbf{Memória global}: é a memória principal da GPU. Pode ser acessada por todas as \textit{threads/cores}, porém possui alta latência e baixo \textit{throughput}.

\item \textbf{Memória compartilhada}: como já mencionado, é a memória dedicada de cada SM que possui baixa latência. Somente \textit{threads} de um mesmo bloco pode acessá-la.

\item \textbf{Memória local}: possui este nome pois é a memória específica de uma \textit{thread}.

\end{itemize}

\begin{figure}[h]
\centering
\includegraphics[scale=0.3]{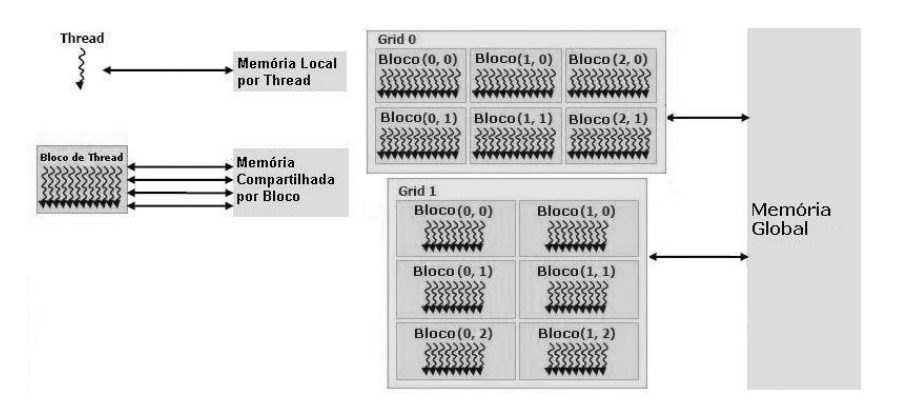}
\caption{Distribuição de \textit{threads} em uma GPU}
\label{fig:threads1}
\end{figure}

\noindent Das memórias descritas, a CPU tem acesso somente a memória global. A distribuição de \textit{threads} e hierárquia de memória é ilustrada pela Figura \ref{fig:threads1}.

Independente da arquitetura da GPU e número de \textit{cores} por SMs, cada \textit{warp} pode abrigar no máximo 32 \textit{threads} executando ao mesmo tempo. Dessa forma para uma melhor performance é recomendado utilizar o número de \textit{threads} por bloco multiplo de 32. Quando um bloco de \textit{threads} é entregue a um SM, o mesmo particiona essas \textit{threads} em \textit{warps} e cada \textit{thread} recebe um índice único. A \textit{thread} decide em qual dado atuar de acordo com seu índice. Por fim, a SM escalona cada bloco de \textit{threads} para execução em seus \textit{cores} e os mesmos são executados de maneira aleatória.

\subsection{Plataforma CUDA}
Como já mencionado, CUDA é a plataforma que permite utilizar uma GPU com propósito geral de computação. CUDA é um \textit{software} desenvolvido pela NVIDIA \textit{corporation}, portanto somente GPUs da NVIDIA são capaz de executar códigos CUDA. Uma opção à plataforma é a OpenCL \cite{AMDzone} que permite codificar tanto para GPUs NVIDIA quanto para ATI/AMD. O desenvolvimento de códigos em CUDA podem ser realizados utilizando C/C++ juntamente com alguns comandos específicos da plataforma. De maneira geral, programadores com alguma experiência em C/C++ não encontram dificuldades com a linguagem.

Para executar uma aplicação em uma GPU sempre será necessário um código na CPU. Basicamente, a codificação consistem em um programa na CPU enviar dados para GPU computar alguma operação específica e a GPU devolve para a CPU o resultado final. Essa comunicação deve ser realizada através da memória principal da CPU e a memória global da GPU. Resumidamente, para se executar um código na GPU são necessários cumprir os seguintes passos em CUDA:

\begin{itemize}
\item Implementar uma função especial chamada \textit{Kernel}. Essa função será executada dentro da GPU e utilizará os índices das \textit{threads} para operar nos dados.

\item Definir a quantidade de grids e blocos definindo assim o número total de \textit{threads} que serão executados na GPU. Esses parâmetro dependem da arquitetura da placa e quanto mais o programador é familiar com a mesma, mais proveito ele tira da plataforma.

\item Enviar os dados a serem executados da memória da CPU para memória global da GPU. Após a conclusão da execução da GPU o caminho inverso deve ser realizado.
\end{itemize}

\noindent Para maior compreensão da plataforma CUDA, bem como detalhes específicos de codificação, sugere-se ao leitor o manual da plataforma \cite{manualCUDA}.

%% file: src/mult-mat.tex
\begin{figure}[h]
\centering
\includegraphics[scale=0.2]{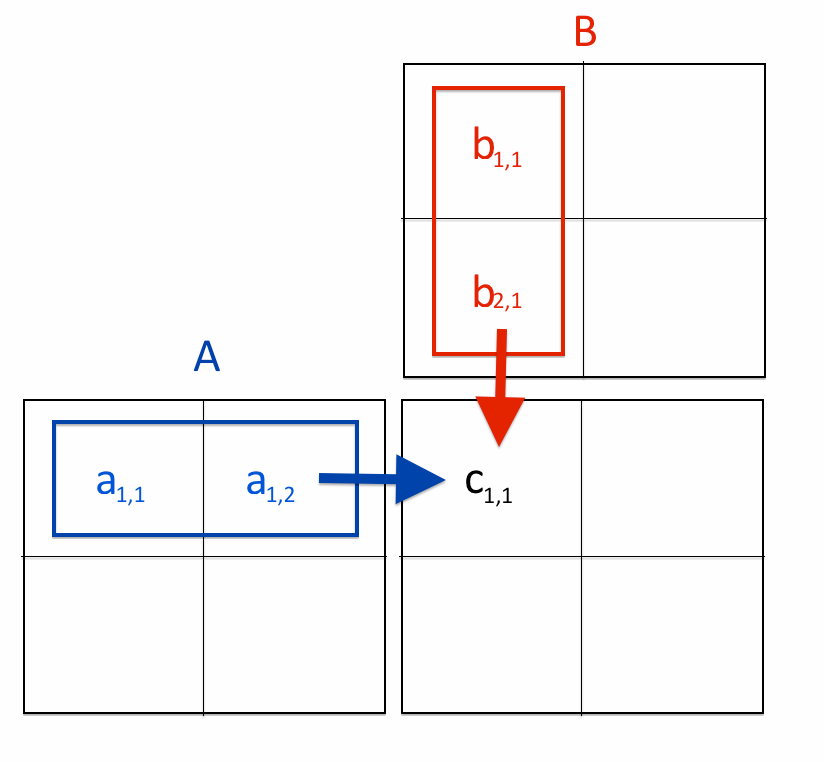}
\caption{Exemplo de multiplicação de matrizes com uma matriz 2 x 2}
\label{fig:exmatmul}
\end{figure}

Nesta seção, é realizada uma discussão da metodologia de multiplicação de matrizes utilizado GPU e OpenMP com CPU. Recapitulando brevemente a multiplicação de matrizes, supõe-se que deseja-se multiplicar as matrizes $A_{m \times n}$ por $B_{n \times w}$. O resultado dessa operação sera uma matriz $C_{m \times w}$, ou seja, $A \times B = C$. A Figura \ref{fig:exmatmul} mostra de maneira intuitiva a ideia por trás da multiplicação de matrizes. Para se obter os elementos de $C$, cada linha de $A$ é multiplicada elemento a elemento por uma coluna de $B$. Ao final o valor é agregado por meio de um somatório. É importante ressaltar que para a multiplicação ser viável, o número de colunas de $A$ deve ser igual ao número de linhas de $B$.

\noindent Dessa maneira é fácil observar que multiplicar matrizes é um bom exemplo de computação paralela. Cada elemento de $C$ é computado de maneira independente, logo, pode ser paralelizado.

\subsection{Multiplicando matrizes na GPU}
\begin{figure}[h]
\centering
\includegraphics[scale=0.45]{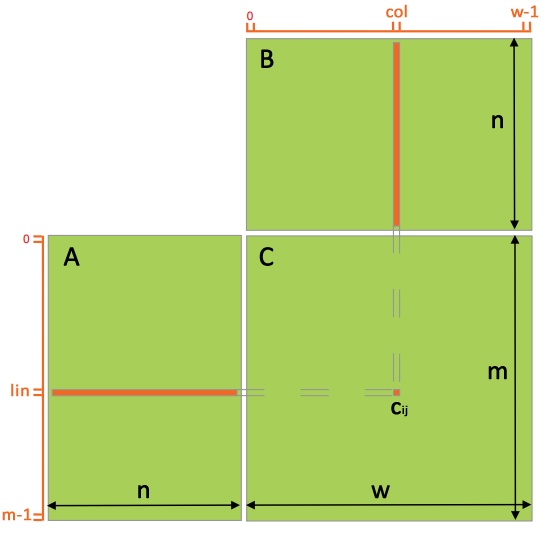}
\caption{Multiplicação de matrizes na qual cada \textit{thread} calcula $c_{ij}$} 
\label{fig:multabs1}
\end{figure}

A primeira abordagem para paralelização da multiplicação de matrizes utilizando uma GPU é disparar diversas \textit{threads} fazendo com que cada uma calcule um elemento da matriz resultante $C$. Nesta abordagem, cada \textit{thread} ler uma linha de $A$ e uma coluna de $B$ para computar o elemento $c_{ij}$ de $C$, sendo $i = 1 \cdots  m$ e $j = 1 \cdots  w$. Na Figura \ref{fig:multabs1} é ilustrada essa abordagem \cite{manualCUDA}. Utilizando essa abordagem de multiplicação, as matrizes $A$ e $B$ são é carregadas na memória global $w$ e $m$ vezes, respectivamente. Com isso o algoritmo faz muitos acessos à memória global, que possui alta latência e baixo \textit{throughput}. 

\begin{figure}[h]
\centering
\includegraphics[scale=0.45]{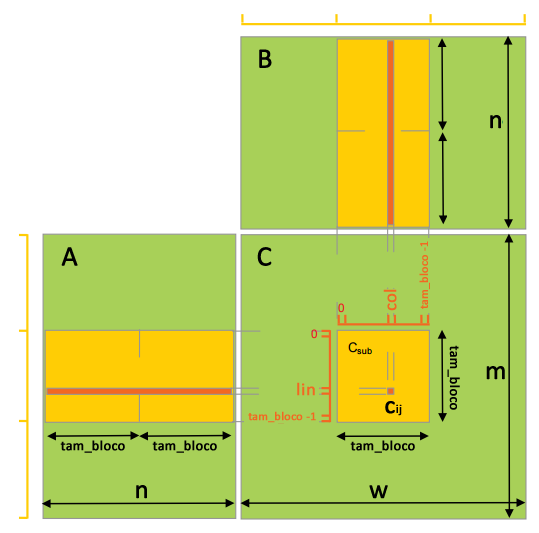}
\caption{Multiplicação de matrizes utilizando ladrilhamento} 
\label{fig:multabs2}
\end{figure}

Com intuito de extrair a eficiência máxima que uma GPU pode entregar, a segunda abordagem de multiplicação de matrizes na GPU tem como objetivo utilizar a memória compartilhada de cada bloco de \textit{threads} visando reduzir o número de acessos a memória global do dispositivo. Para isso as matrizes são subdivididas em pequenos blocos como mostrado na Figura \ref{fig:multabs2} \cite{manualCUDA}. Essa técnica é conhecida como multiplicação por ladrilhamento e diferentemente da abordagem anterior, no qual toda linha de $A$ e coluna $B$ era multiplicada de uma só vez gerando assim um elemento de $C$, no ladrilhamento tem-se submatrizes de $A$ e $B$ que vão gerar um valor parcial de um elemento de $C$. Quando todos os ladrilhos forem processados o valor final de cada um dos elementos da matriz $C$ será o somatório de cada elemento parcial obtido pela multiplicação de cada uma das linhas e colunas submatrizes $A$ e $B$, respectivamente.

O objetivo de principal de se usar o ladrilhamento em GPU é carregar cada submatriz na memória compartilhada do bloco de \textit{threads}. O caso ideal seria alocar toda matriz dentro de uma memória compartilhada. Porém, como já mencionado, essa memória é de tamanho reduzido quando comparada com a memória global. Dessa forma, a ideia é carregar na memória compartilhada apenas submatrizes de $A$ e $B$, na qual cada bloco de \textit{threads} pode compartilhar seus dados de maneira rápida. Com isso, um requisito de projeto é que o ladrilho caiba dentro de um bloco de \textit{threads} da GPU. O valor do tamanho do bloco varia de acordo com o dispositivo e cabe ao programador conhecer a arquitetura para melhor aproveitá-la.

Fazendo uma breve comparação entre as duas abordagens de multiplicação de matrizes em GPU apresentadas anteriormente, a Figura \ref{fig:complad} apresenta um gráfico de tempo de execução de cada um dos métodos em relação a ordem das matrizes multiplicadas. É possível observar o ganho de performance na medida que as matrizes aumentam. A arquitetura utilizada para gerar este gráfico será discutida na seção IV. 

\begin{figure}[h]
\centering
\includegraphics[scale=0.6]{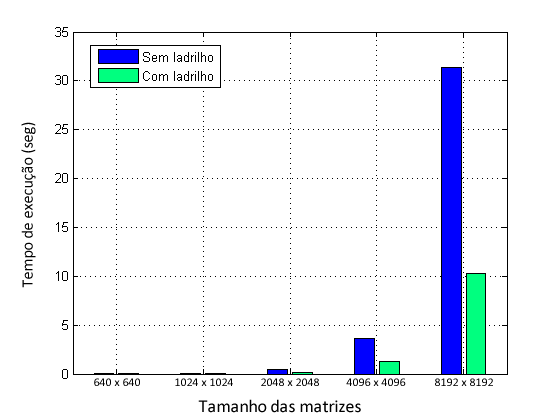}
\label{fig:complad}
\caption{Comparação entre as abordagens com e sem ladrilhamento na GPU utilizando precisão simples. As matrizes A e B são quadradas e os tamanhos são indicados no gráfico} 
\label{fig:complad}
\end{figure}

\subsection{Multiplicando matrizes na CPU com OpenMP}
\begin{figure}[h]
\centering
\includegraphics[scale=0.6]{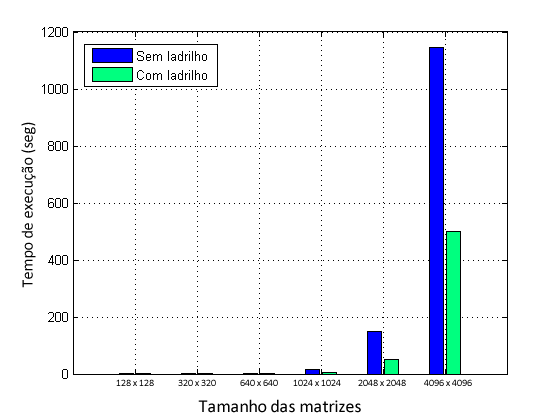}
\caption{Comparação entre as abordagens com e sem ladrilhamento na CPU para precisão simples. As matrizes A e B são quadradas e os tamanhos são indicados no gráfico}
\label{fig:complad2}
\end{figure}

A ideia de ladrilhamento das matrizes, discutida na seção anterior, também aplicada na CPU. Na GPU, as \textit{threads} são disparadas e com os índices de cada uma delas é possível operar sobre os dados de acordo com seus blocos. Na CPU esses índices são obtidos por loops de controle. O objetivo principal do ladrilhamento também é o mesmo: tirar proveito da arquitetura de CPU através da memória cache de acesso rápido. Mesmo no código sequencial já se obtém um ganho de performance, como mostrado na Figura \ref{fig:complad2}

Para realizar paralelização por meio da CPU é utilizado a OpenMP. A API dá suporte as linguagens C/C++ e Fortran, e basicamente o que deve ser incluído ao código original são diretiva em trechos de códigos que devem ser paralelizado. Com isso, a API é capaz de disparar \textit{threads} utilizando os núcleos da CPU em questão. Um exemplo de diretiva para paralelizar um loop em C é descrito a seguir:

\begin{verbatim}
#pragma omp parallel for default(none) \
  shared(n,x,y) private(i)
for (i=0; i<n; i++)
    	x[i] += y[i];
}
\end{verbatim}

\noindent A simples anotação indica para paralelizar o loop na sequência, compartilhar as variáveis $n, x$ e $y$ e manter $i$ privado. Dessa maneira simples já é possível obter ganho de desempenho utilizando os núcles da CPU. O número de \textit{threads} disparadas como \textit{default} é uma por núcleo do processador, todavia o programador pode alterar esse valor por meio da variável de ambiente OMP\_NUM\_THREAD. Para mais informações sobre OpenMP sugere-se \cite{chapman2008}. Na próxima seção será abordada a comparação entre a implementação sequencial e paralela utilizando OpenMP.

%% file: src/experimentos.tex
Nesta seção são realizados experimentos para comparar a performance da multiplicação de matrizes utilizando a computação sequencial e a paralela utilizando CUDA e OpenMP. Os experimentos foram executados em uma máquina com sistema operacional Linux, distribuição Xubuntu, com processador intel core i7, 2.5 GHz, 2 núcleos, 3 MB de memória cache e 6GB de memória RAM e uma placa gráfica NVIDIA Geforce 940M. A apresentação da GPU e as considerações para máximo desempenho serão descritos na sequências. O código de todos os experimentos esta disponível neste \href{https://github.com/paaatcha/TrabGPU/tree/master/Trab1}{repositório do Github}.

Os experimentos serão realizados da seguinte forma:
\begin{itemize}
\item A multiplicação será realizada considerando a implementação por ladrilho sequencial, paralelizada em OpenMP e em GPU
\item As abordagens serão aplicadas para precisão simples (\textit{float}) e dupla (\textit{double})
\item A multiplicação será executada em três diferentes configurações: 1 vez, 100 vezes, e 1000 vezes. As duas últimas paralelizadas com OpenMP
\item As matrizes e os ladrilhos serão ajustados para máximo desempenho
\item O desempenho será medido em termos de tempo de execução, \textit{speedup}, através da lei de Amdahl \cite{hennessy2011}, e em GFLOPS
\end{itemize}

\subsection{Características da GPU utilizada}
A GPU utilizada neste trabalho, modelo Geforce 940M, possui as seguintes características:
\begin{itemize}
\item 3 SMs
\item 384 CUDA \textit{cores}, 128 por SM
\item 32 \textit{threads} por \textit{warp}
\item Máximo de 1024 \textit{threads} por bloco
\item Máximo de 2048 \textit{threads} SM
\item Máximo de 32 blocos por SM
\item Memória global: 2 GB
\item Memória compartilhada: 49 KB
\item Processamento máximo pode atingir 790.3 GFLOPS para precisão simples e 24.7 GFLOPS para dupla

\end{itemize}

\begin{figure}[h]
\centering
\includegraphics[scale=0.6]{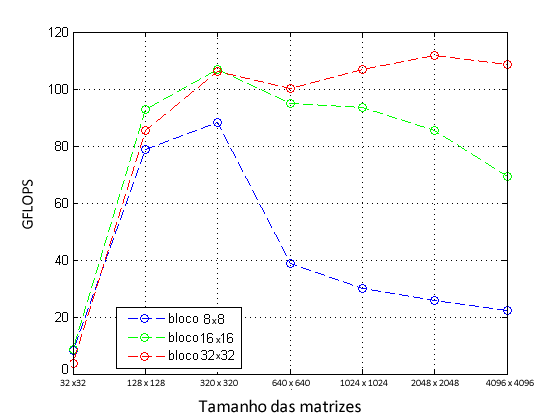}
\caption{Comparação de GFLOPS de acordo com tamanho do bloco para precisão simples. As matrizes A e B são quadradas e os tamanhos são idicados no gráfico} 
\label{fig:comp_blocos}
\end{figure}

\noindent De acordo com as configurações do dispositivo, o tamanho máximo do ladrilho, para que o mesmo encaixe em um bloco, deverá ser $32 \times 32 = 1024$ \textit{threads}, limite da placa. Como a SM suporta até 2048 \textit{threads}, o número de blocos por SM será igual a $2048/1024 = 2$ blocos, respeitando também o limite de blocos por SM. Na Figura \ref{fig:comp_blocos} é ilustrado a variação de GFLOPS de acordo com o tamanho do bloco escolhido para dados de precisão simples. Pode ser observado que na medida que as matrizes aumentam, as configurações de blocos $8 \times 8$ e $16 \times 16$ diminuem a capacidade de cálculo em relação ao tamanho $32 \times 32$. 

\subsection{Experimento parte I - Multiplicação para precisão simples}

\begin{figure}[h]
\centering
\includegraphics[scale=0.6]{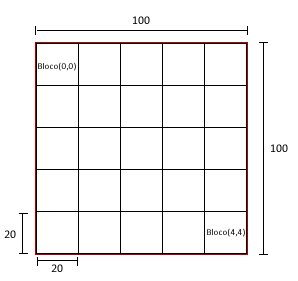}
\caption{Um grid de blocos $5 \times 5$ em uma matriz $100 \times 100$} 
\label{fig:grid}
\end{figure}

Nesta primeira etapa será analisado os resultados para matrizes de precisão simples (\textit{float}). Para obter eficiência máxima, as matrizes escolhidas possuem número de linhas e colunas divisíveis pelo tamanho máximo de ladrilho, ou seja um bloco de  $32 \times 32$. Dessa forma um grid de blocos se encaixa perfeitamente nas dimensões da matriz, sem necessidade de verificações na função de kernel da GPU, o que ocorreria se o ladrilho não fosse divisível. Na Figura \ref{fig:grid} é ilustrado um exemplo simples de distribuição de blocos na matriz. Neste caso existe uma matriz $100 \times 100$ e um ladrilho/bloco $20 \times 20$, como a divisão dimensões da matriz pelo bloco é inteira, é utilizado um grid de blocos de $5 \times 5$.

Como o intuito deste trabalho é simplesmente comparar performance, essa premissa é compreensível. Que fique claro que isso não é uma limitação da GPU, é simplesmente um artifício para verificar performance máxima. Nada impede da verificação ser acrescida no Kernel e o tamanho da matriz ser variável. As ordens das matrizes escolhidas, tanto de $A$ quanto de $B$, variam de $32 \times 32$ até $2048 \times 2048$. Com esses valores, tanto para precisão simples, quanto para precisão dupla, as matrizes cabem dentro da memória global. No caso máximo, considerando a precisão dupla, $2048$ bytes $\times$ $2048$ bytes $\times$ $8$ bytes $= 32$ MB, ao alocar espaço para 3 matrizes deste mesmo tamanho, tem-se $96$ MB, bem abaixo dos 2GB possíveis.

Para a abordagem sequencial e paralela via OpenMP o tamanho do bloco escolhido é o mesmo da GPU. Além disso, como a CPU utilizada possui apenas dois núcleos, o número de \textit{threads} disparadas será igual a quatro, duas por núcleo. Para facilitar a visualização dos gráficos, primeiramente a abordagem sequencial é comparada com a paralela via OpenMP. Na sequencia, a abordagem paralela via OpenMP é comparada com a GPU. Essa divisão foi escolhida pois, conforme será exposto, o desempenho da GPU é extremamente melhor do que as duas anteriores. Se todos os gráficos fossem plotados juntos seria muito difícil comparar as duas piores por questão de escala.

\hfill
\subsubsection{CPU $\times$ CPU + OpenMP}
\hfill

\begin{figure*}
\centering    
\subfigure[Desempenho em GFLOPS]{
\includegraphics[width=0.6\columnwidth]{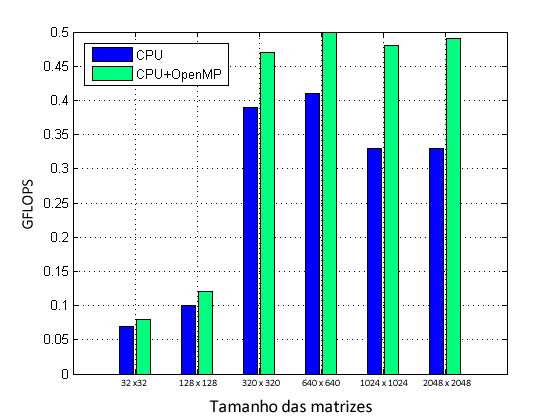}
\label{fig:cpu_simples_1a}
}
\quad
\subfigure[Desempenho em tempo de execução]{
\includegraphics[width=0.6\columnwidth]{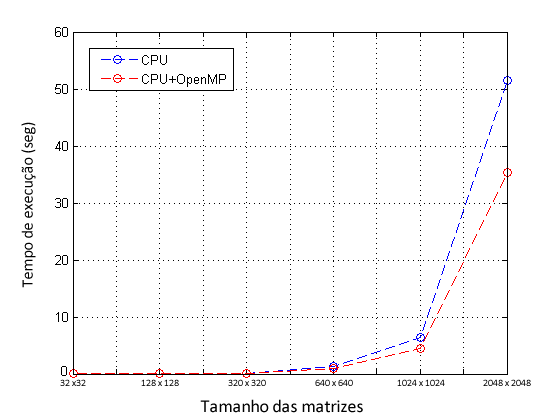}
\label{fig:cpu_simples_1b}
}
\quad
\subfigure[\textit{Speedup} da CPU + OpenMP em relação a CPU]{
\includegraphics[width=0.6\columnwidth]{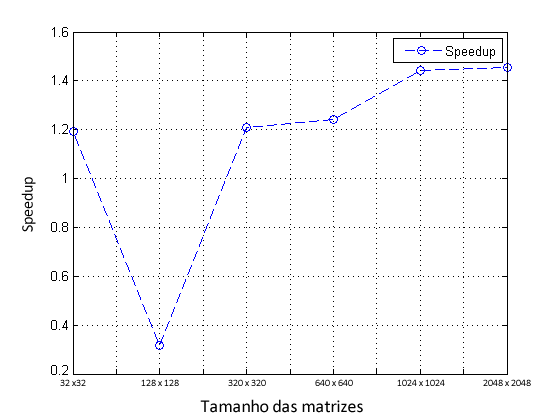}
\label{fig:cpu_simples_1c}
}

\caption{Desempenho CPU $\times$ CPU + OpenMP para multiplicação de matrizes executada apenas uma vez - Precisão simples}
\label{fig:cpu_simples_1}
\end{figure*}

O desempenho do experimento executado apenas uma vez utilizando a CPU e a CPU + OpenMP é ilustrado nos gráficos da Figura \ref{fig:cpu_simples_1}. Como mostrado na Figura \ref{fig:cpu_simples_1a}, a partir da multiplicação das matrizes de ordem $320 \times 320$ já se obtém um pequeno ganho de de processamento. O tempo de execução, como ilustrado na Figura \ref{fig:cpu_simples_1b}, começa a fazer diferença na ordem $1024 \times 1024$, o que também é mostrado no gráfico de \textit{speedup} na \ref{fig:cpu_simples_1c}. 

\begin{figure*}
\centering    
\subfigure[Desempenho em GFLOPS]{
\label{fig:cpu_simples_2a}
\includegraphics[width=0.6\columnwidth]{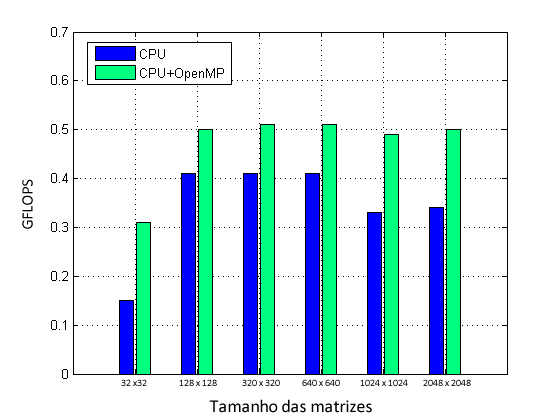}
}
\quad
\subfigure[Desempenho em tempo de execução]{
\label{fig:cpu_simples_2b}
\includegraphics[width=0.6\columnwidth]{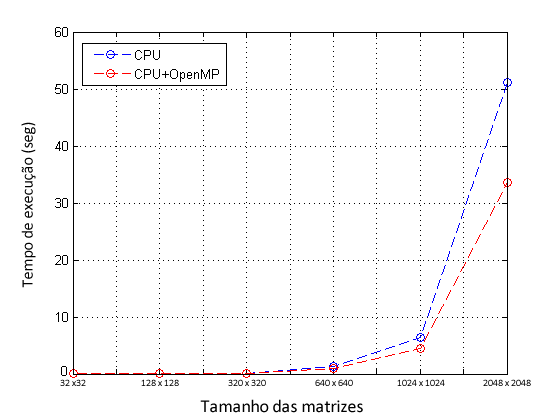}
}
\quad
\subfigure[\textit{Speedup} da CPU + OpenMP em relação a CPU]{
\label{fig:cpu_simples_2c}
\includegraphics[width=0.6\columnwidth]{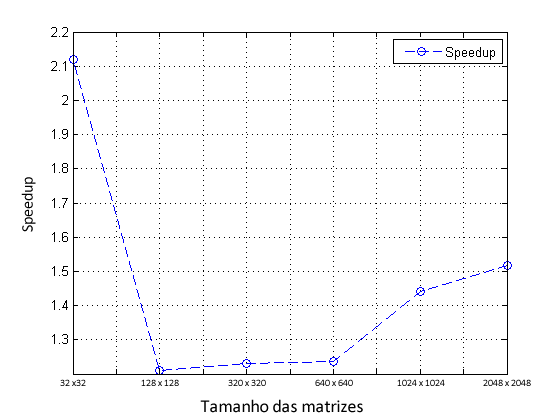}
}

\caption{Desempenho CPU $\times$ CPU + OpenMP para multiplicação de matrizes executada 100 vezes - Precisão simples}
\label{fig:cpu_simples_2}
\end{figure*}

\begin{figure*}
\centering    
\subfigure[Desempenho em GFLOPS]{
\label{fig:cpu_simples_3a}
\includegraphics[width=0.6\columnwidth]{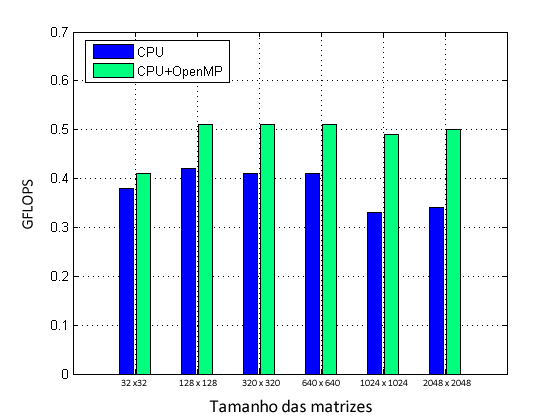}
}
\quad
\subfigure[Desempenho em tempo de execução]{
\label{fig:cpu_simples_3b}
\includegraphics[width=0.6\columnwidth]{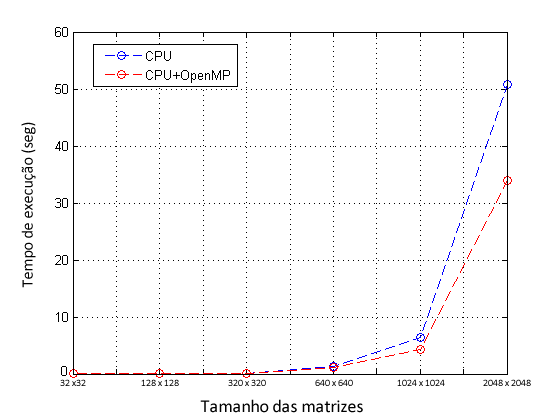}
}
\quad
\subfigure[\textit{Speedup} da CPU + OpenMP em relação a CPU]{
\label{fig:cpu_simples_3c}
\includegraphics[width=0.6\columnwidth]{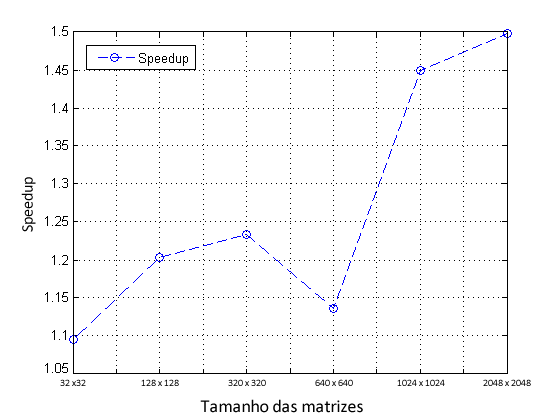}
}

\caption{Desempenho CPU $\times$ CPU + OpenMP para multiplicação de matrizes executada 1000 vezes - Precisão simples}
\label{fig:cpu_simples_3}
\end{figure*}

Comparando o desempenho anterior com as execuções de 100 e 1000 vezes, ilustradas nos gráficos da Figura \ref{fig:cpu_simples_2} e \ref{fig:cpu_simples_3}, respectivamente, é possível notar nos gráficos das Figura \ref{fig:cpu_simples_2a} e \ref{fig:cpu_simples_3a} que o processamento cresce um pouco, principalmente para as multiplicações com matrizes de menor ordem. A diferença de tempo computacional, apontada nos gráficos das Figura \ref{fig:cpu_simples_2b} e \ref{fig:cpu_simples_3b} se mantém praticamente a mesma. Por fim ocorre uma variação de \textit{speedup}, quando comparado os gráficos os gráficos das Figura \ref{fig:cpu_simples_1c}, \ref{fig:cpu_simples_2c} e \ref{fig:cpu_simples_3c}. Neste caso, como os tempos de execução até a ordem $320 \times 320$ são bem baixos (na ordem de mseg), o valor do \textit{speedup} é mais constante e coerente a partir da ordem $640 \times 640$, se mantendo acima de 1, com máximo de 1.5.

De maneira geral a paralelização via OpenMP obtém um ganho em relação a sequencial quando as matrizes multiplicadas alcançam a ordem de $1024 \times 1024$. Utilizando essa CPU o ganho não tão evidente pois a mesma possui apenas dois núcleos, o que limita o potencial do \textit{framework}. Todavia, devido a facilidade de se incluir a paralelização, ainda assim é recomendável o uso da API para matrizes grandes.

\subsubsection{CPU + OpenMP $\times$ GPU}

\begin{figure*}
\centering    
\subfigure[Desempenho em GFLOPS]{
\label{fig:gpu_simples_1a}
\includegraphics[width=0.6\columnwidth]{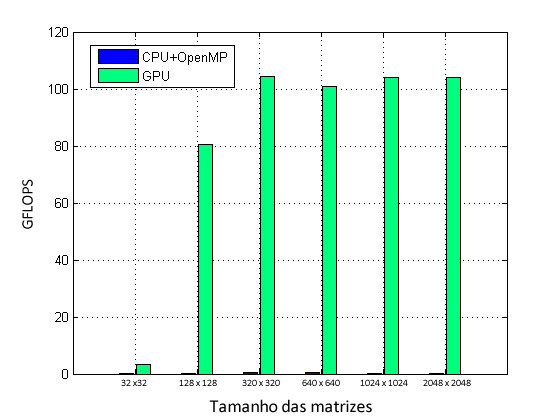}
}
\quad
\subfigure[Desempenho em tempo de execução]{
\label{fig:gpu_simples_1b}
\includegraphics[width=0.6\columnwidth]{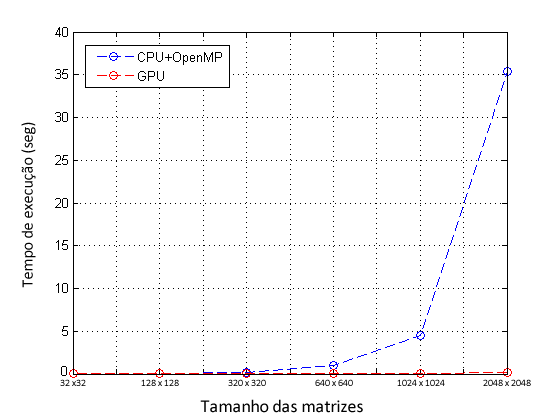}
}
\quad
\subfigure[\textit{Speedup} da CPU + OpenMP em relação a CPU]{
\label{fig:gpu_simples_1c}
\includegraphics[width=0.6\columnwidth]{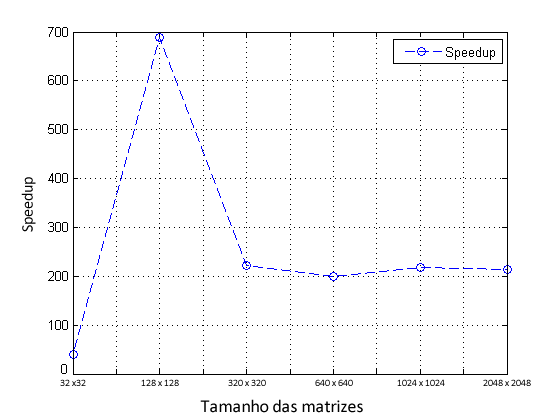}
}

\caption{Desempenho GPU $\times$ CPU + OpenMP para multiplicação de matrizes executada apenas uma vez - Precisão simples}
\label{fig:gpu_simples_1}
\end{figure*}

O desempenho do experimento executado apenas uma vez é ilustrado nos gráficos da Figura \ref{fig:gpu_simples_1}. Como mostrado na Figura \ref{fig:gpu_simples_1a}, na primeira configuração de matriz a GPU possui uma taxa de GFLOPS cerca de 30x maior do que a CPU+OpenMP. A partir desta configuração a proporção dispara para mais de 200x, deixando claro que a GPU efetua muito mais cálculos por segundo do que a CPU e a CPU+OpenMP. A diferença do tempo de execução, até a multiplicação de matrizes de $640 \times 640$, não é tão perceptível pois está na ordem de milissegundos, como ilustrado na Figura \ref{fig:gpu_simples_1b}. A partir da configuração $1024 \times 1024$ o tempo de execução começa a fazer bastante diferença, sendo que para última configuração a GPU executa a multiplicação em torno de 1 seg e a CPU+OpenMP mais de 50 seg. Por fim, o \textit{speedup} é ilustrado na Figura \ref{fig:gpu_simples_1c}, na qual é possível observar a superioridade da GPU para com a CPU+OpenMP, principalmente quando a ordem das matrizes aumentam. 
        
\begin{figure*}
\centering    
\subfigure[Desempenho em GFLOPS]{
\label{fig:gpu_simples_2a}
\includegraphics[width=0.6\columnwidth]{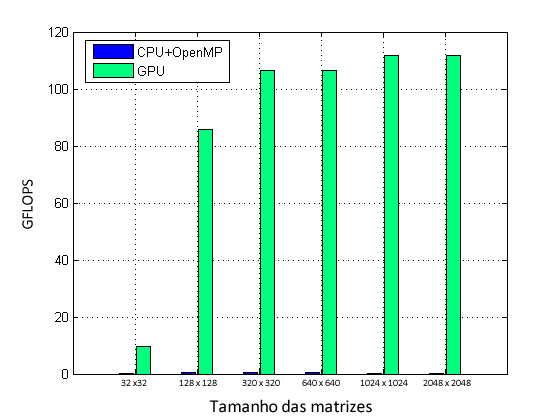}
}
\quad
\subfigure[Desempenho em tempo de execução]{
\label{fig:gpu_simples_2b}
\includegraphics[width=0.6\columnwidth]{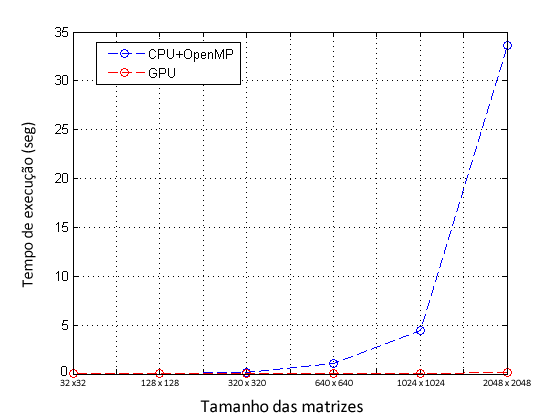}
}
\quad
\subfigure[\textit{Speedup} da CPU + OpenMP em relação a CPU]{
\label{fig:gpu_simples_2c}
\includegraphics[width=0.6\columnwidth]{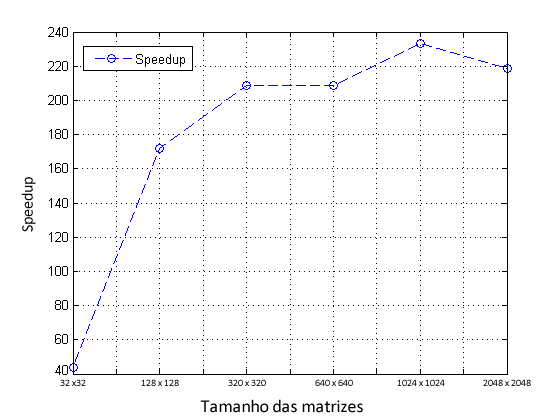}
}

\caption{Desempenho GPU $\times$ CPU + OpenMP para multiplicação de matrizes executada 100 vezes - Precisão simples}
\label{fig:gpu_simples_2}
\end{figure*}

\begin{figure*}
\centering    
\subfigure[Desempenho em GFLOPS]{
\label{fig:gpu_simples_3a}
\includegraphics[width=0.6\columnwidth]{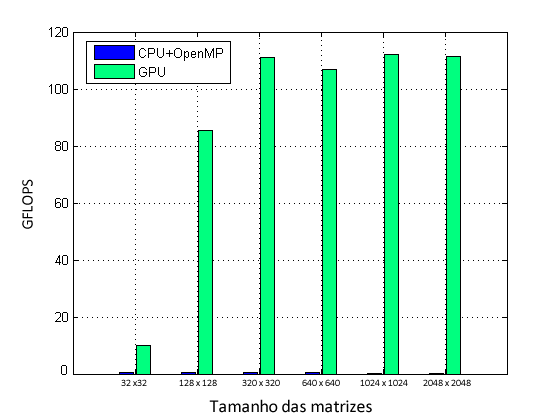}
}
\quad
\subfigure[Desempenho em tempo de execução]{
\label{fig:gpu_simples_3b}
\includegraphics[width=0.6\columnwidth]{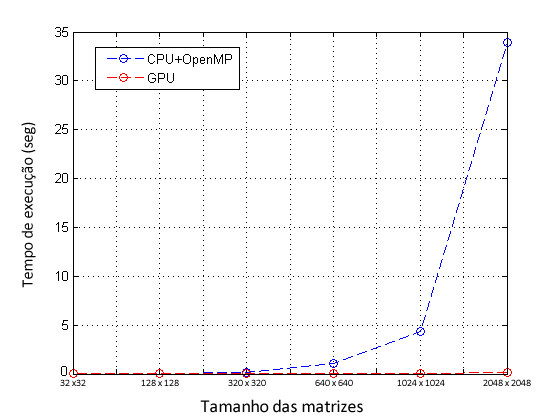}
}
\quad
\subfigure[\textit{Speedup} da CPU + OpenMP em relação a CPU]{
\label{fig:gpu_simples_3c}
\includegraphics[width=0.6\columnwidth]{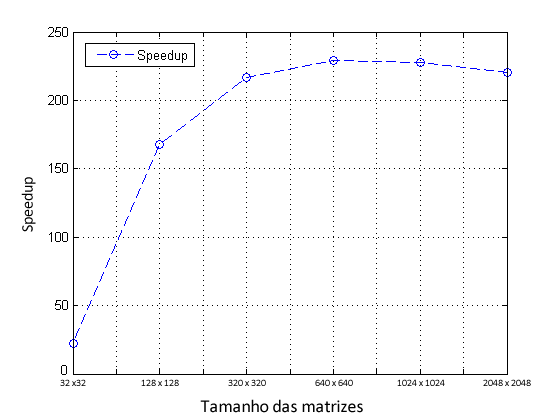}
}

\caption{Desempenho GPU $\times$ CPU + OpenMP para multiplicação de matrizes executada 1000 vezes - Precisão simples}
\label{fig:gpu_simples_3}
\end{figure*}

O desempenho dos experimentos executando a multiplicação 100 e 1000 vezes são ilustrados nas Figura \ref{fig:gpu_simples_2} e \ref{fig:gpu_simples_3}, respectivamente. Sendo assim, é possível observar que a diferença de GFLOPS se mantém. Todavia, no caso da GPU, o valor aumenta um pouco para as matrizes menores. No caso da ordem $32 \times 32$ ocorre um acréscimo de quase 3x. As diferenças de tempo de execução também se matém, oscilando muito pouco. Com isso o valor do \textit{speedup} também oscila pouco, sendo o valor mínimo em torno de 25 e o máximo por volta de 325.

De maneira geral, baseado nos gráficos apresentados nas Figuras \ref{fig:cpu_simples_1} a \ref{fig:gpu_simples_3}, é possível notar a diferença de desempenho da GPU para CPU e CPU+OpenMP. A capacidade de processamento é extremamente maior sendo muito vantajoso seu uso, principalmente quando a multiplicação de matrizes atinge a ordem de $1024 \times 1024$, no qual o crescimento da curva de tempo de execução para CPU e CPU+OpenMP é muito alto.

\subsection{Experimento parte II - Multiplicação para precisão dupla}
Nesta segunda parte de experimentos serão analisados os resultados para matrizes de precisão dupla (\textit{double}). As configurações de matrizes e testes seguem os mesmos moldes descritos na parte I. Nesta ocasião é esperado um queda do poder de processamento da GPU e consequentemente o aumento no tempo de execução da matriz de \textit{doubles}. Isso ocorre devido ao fato da GPU ter menos processadores dedicados a cálculo de precisão dupla.

\subsubsection{CPU $\times$ CPU + OpenMP}

\begin{figure*}
\centering    
\subfigure[Desempenho em GFLOPS]{
\label{fig:cpu_duplo_1a}
\includegraphics[width=0.6\columnwidth]{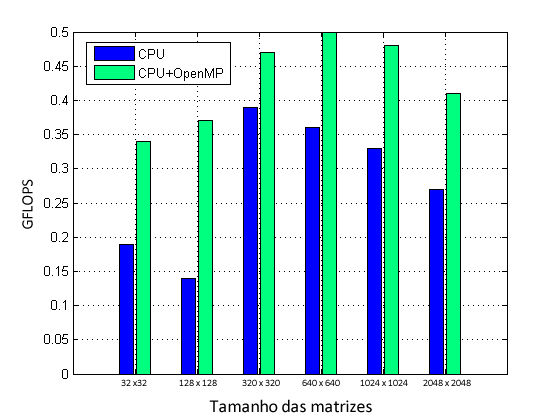}
}
\quad
\subfigure[Desempenho em tempo de execução]{
\label{fig:cpu_duplo_1b}
\includegraphics[width=0.6\columnwidth]{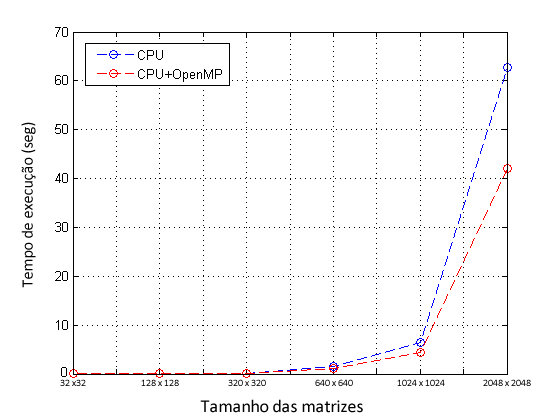}
}
\quad
\subfigure[\textit{Speedup} da CPU + OpenMP em relação a CPU]{
\label{fig:cpu_duplo_1c}
\includegraphics[width=0.6\columnwidth]{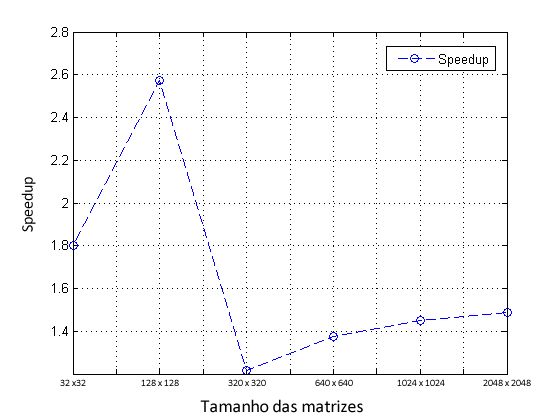}
}

\caption{Desempenho CPU $\times$ CPU + OpenMP para multiplicação de matrizes executada apenas uma vez - Precisão Dupla}
\label{fig:cpu_duplo_1}
\end{figure*}

O desempenho do experimento executado apenas uma vez utilizando a CPU e a CPU + OpenMP para precisão dupla é ilustrado nos gráficos da Figura \ref{fig:cpu_duplo_1}. Como mostrado na Figura \ref{fig:cpu_duplo_1a}, o processamento é bem próximo ao da precisão simples. A diferença de tempo de execução também segue a mesma linha, sendo perceptível a partir da ordem $640 \times 640$, como mostrado na Figura \ref{fig:cpu_duplo_1b}. Como os tempos até a ordem $320 \times 320$ estão na ordem de milissegundos, o \textit{speedup} e mais constante a partir da ordem $640 \times 640$, como mostrado na Figura \ref{fig:cpu_duplo_1c}.

\begin{figure*}
\centering    
\subfigure[Desempenho em GFLOPS]{
\label{fig:cpu_duplo_2a}
\includegraphics[width=0.6\columnwidth]{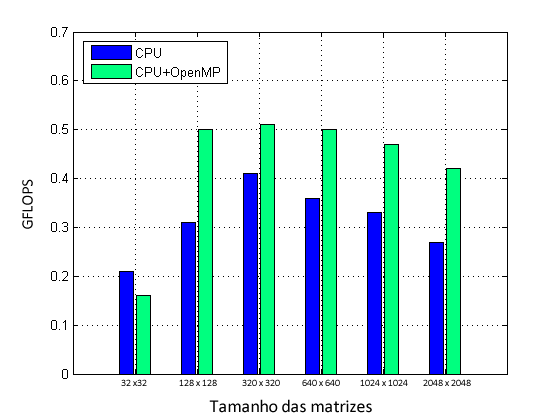}
}
\quad
\subfigure[Desempenho em tempo de execução]{
\label{fig:cpu_duplo_2b}
\includegraphics[width=0.6\columnwidth]{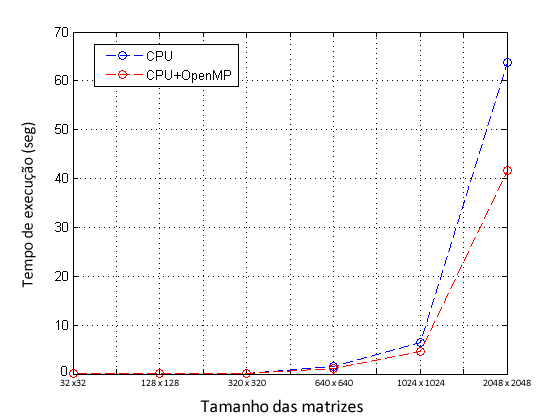}
}
\quad
\subfigure[\textit{Speedup} da CPU + OpenMP em relação a CPU]{
\label{fig:cpu_duplo_2c}
\includegraphics[width=0.6\columnwidth]{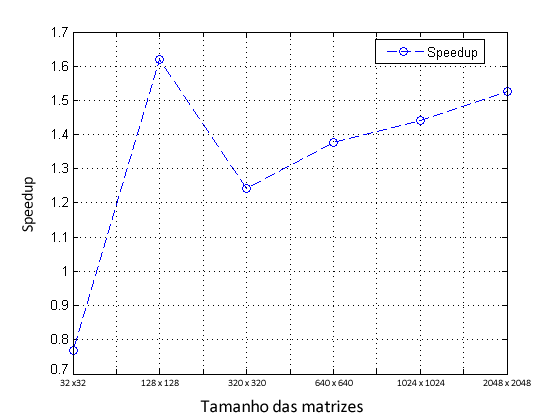}
}

\caption{Desempenho CPU $\times$ CPU + OpenMP para multiplicação de matrizes executada 100 vezes - Precisão Dupla}
\label{fig:cpu_duplo_2}
\end{figure*}

\begin{figure*}
\centering    
\subfigure[Desempenho em GFLOPS]{
\label{fig:cpu_duplo_3a}
\includegraphics[width=0.6\columnwidth]{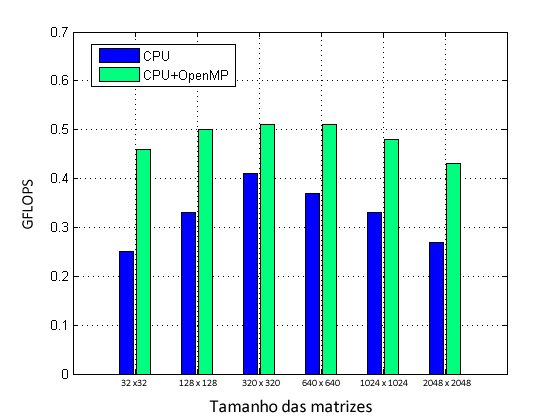}
}
\quad
\subfigure[Desempenho em tempo de execução]{
\label{fig:cpu_duplo_3b}
\includegraphics[width=0.6\columnwidth]{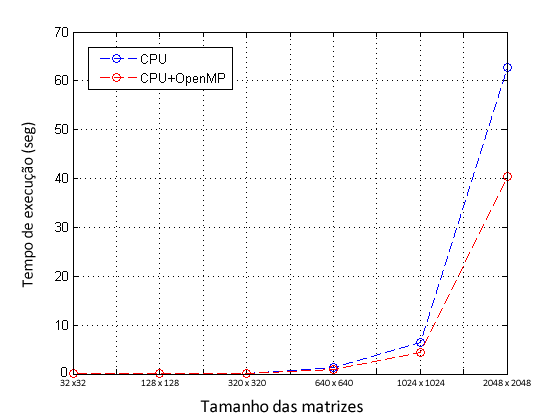}
}
\quad
\subfigure[\textit{Speedup} da CPU + OpenMP em relação a CPU]{
\label{fig:cpu_duplo_3c}
\includegraphics[width=0.6\columnwidth]{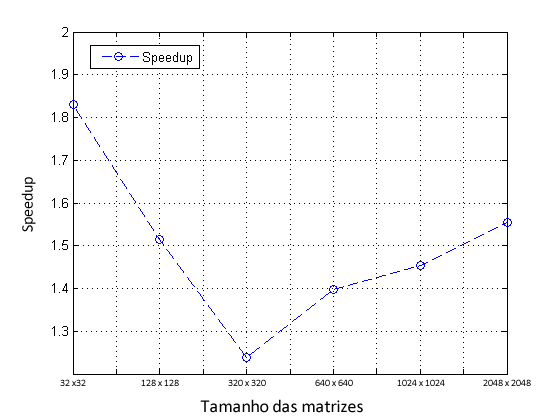}
}

\caption{Desempenho CPU $\times$ CPU + OpenMP para multiplicação de matrizes executada 1000 vezes - Precisão Dupla}
\label{fig:cpu_duplo_3}
\end{figure*}

Novamente, comparando o desempenho anterior com as execuções de 100 e 1000 vezes, ilustradas nos gráficos da Figura \ref{fig:cpu_duplo_2} e \ref{fig:cpu_duplo_3}, respectivamente, é possível notar nos gráficos das Figuras \ref{fig:cpu_duplo_2a} e \ref{fig:cpu_duplo_3a} que o processamento estabiliza em torno de 0.4 e 0.5. A diferença de tempo computacional, apontada nos gráficos das Figuras \ref{fig:cpu_duplo_2b} e \ref{fig:cpu_duplo_3b} sem mantém quase que inalterados. Por fim, o \textit{speedup} possui um comportamento um tanto quanto estranho para as duas primeiras matrizes, todavia, para as demais, o valor estabiliza entre 1.3 e 1.6. 

A conclusão da comparação entre as duas abordagens é semelhante a precisão simples, a paralelização via OpenMP obtém um ganho em relação a sequencial quando as matrizes multiplicadas alcançam a ordem de $1024 \times 1024$. Vale a pena ressaltar novamente, que essa CPU possui apenas 2 núcleos de processamento.

\subsubsection{GPU $\times$ CPU + OpenMP}

\begin{figure*}
\centering    
\subfigure[Desempenho em GFLOPS]{
\label{fig:gpu_duplo_1a}
\includegraphics[width=0.6\columnwidth]{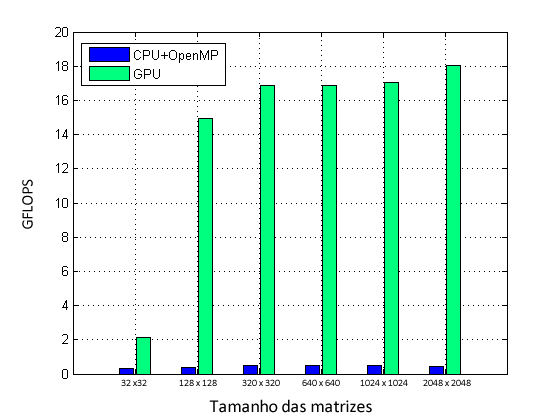}
}
\quad
\subfigure[Desempenho em tempo de execução]{
\label{fig:gpu_duplo_1b}
\includegraphics[width=0.6\columnwidth]{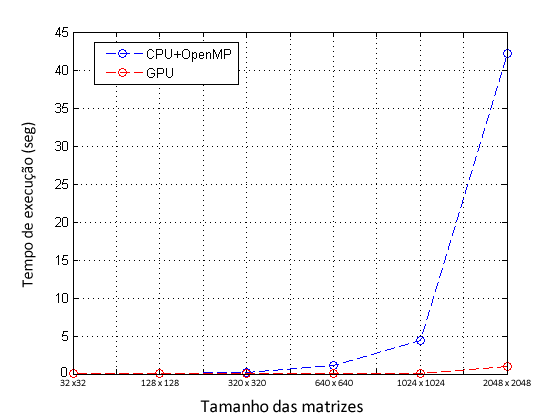}
}
\quad
\subfigure[\textit{Speedup} da CPU + OpenMP em relação a CPU]{
\label{fig:gpu_duplo_1c}
\includegraphics[width=0.6\columnwidth]{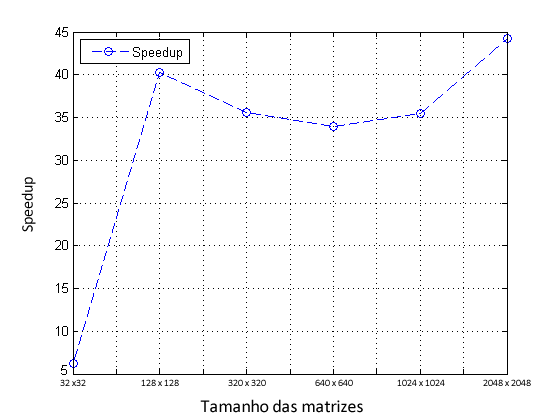}
}

\caption{Desempenho GPU $\times$ CPU + OpenMP para multiplicação de matrizes executada apenas uma vez - Precisão Dupla}
\label{fig:gpu_duplo_1}
\end{figure*}

\begin{figure*}
\centering    
\subfigure[Desempenho em GFLOPS]{
\label{fig:gpu_duplo_2a}
\includegraphics[width=0.6\columnwidth]{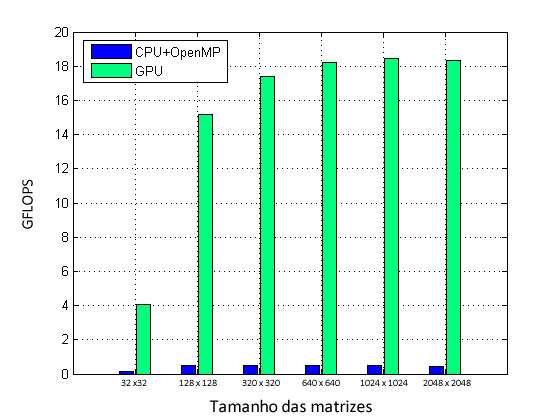}
}
\quad
\subfigure[Desempenho em tempo de execução]{
\label{fig:gpu_duplo_2b}
\includegraphics[width=0.6\columnwidth]{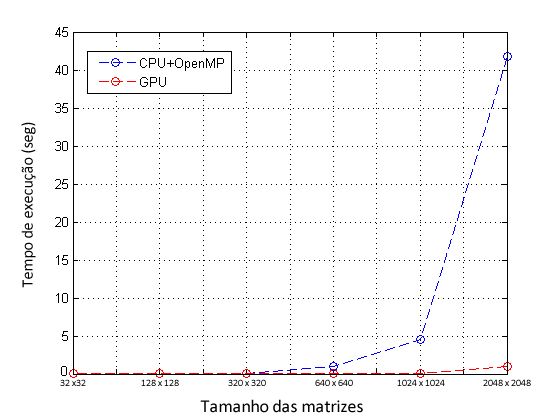}
}
\quad
\subfigure[\textit{Speedup} da CPU + OpenMP em relação a CPU]{
\label{fig:gpu_duplo_2c}
\includegraphics[width=0.6\columnwidth]{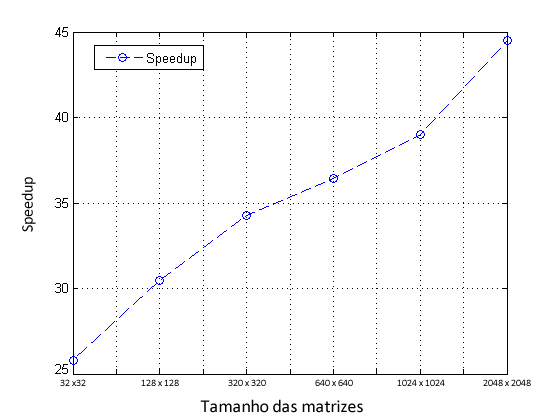}
}

\caption{Desempenho GPU $\times$ CPU + OpenMP para multiplicação de matrizes executada 100 vezes - Precisão Dupla}
\label{fig:gpu_duplo_2}
\end{figure*}

\begin{figure*}
\centering    
\subfigure[Desempenho em GFLOPS]{
\label{fig:gpu_duplo_3a}
\includegraphics[width=0.6\columnwidth]{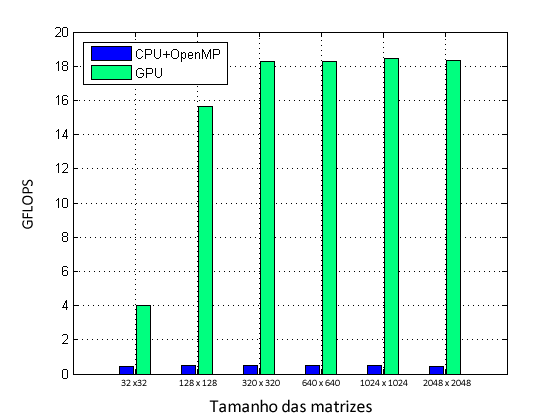}
}
\quad
\subfigure[Desempenho em tempo de execução]{
\label{fig:gpu_duplo_3b}
\includegraphics[width=0.6\columnwidth]{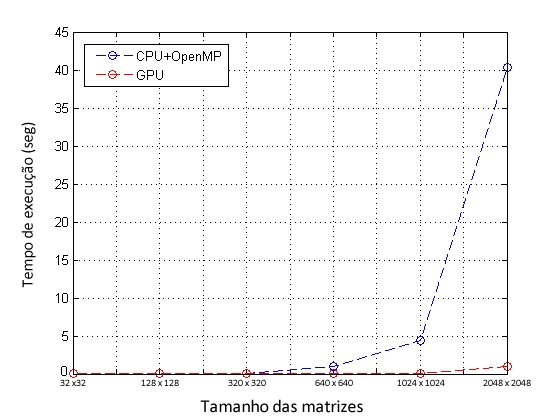}
}
\quad
\subfigure[\textit{Speedup} da CPU + OpenMP em relação a CPU]{
\label{fig:gpu_duplo_3c}
\includegraphics[width=0.6\columnwidth]{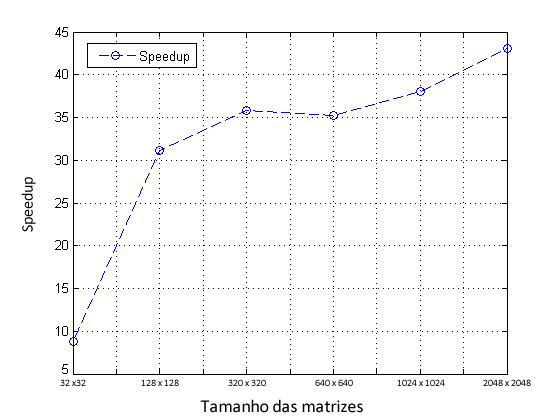}
}

\caption{Desempenho GPU $\times$ CPU + OpenMP para multiplicação de matrizes executada 1000 vezes - Precisão Dupla}
\label{fig:gpu_duplo_3}
\end{figure*}

Desta vez o experimento é executado 1, 100 e 1000 vezes para precisão dupla, como ilustrados nas Figuras \ref{fig:gpu_duplo_1}, \ref{fig:gpu_duplo_2} e \ref{fig:gpu_duplo_3}, respectivamente. Para o caso de uma execução, na Figura \ref{fig:gpu_duplo_1a} é ilustrado a diferença de GFLOPS entre a CPU+OpenMP e a GPU. Ainda existe uma boa diferença de processamento, todavia muito menor do que a apresentada para a precisão simples. A diferença de tempo de execução, mostrado no gráfico da Figura \ref{fig:gpu_duplo_1b}, ainda é bem discrepante quando a ordem das matrizes multiplicadas aumentam. O \textit{speedup}, mostrado na Figura \ref{fig:gpu_duplo_1c}, também diminui bastante em relação a precisão simples, porém ainda é possível observar o ganho em relação a GPU. Especificamente para a multiplicação de matrizes de ordem $128 \times 128$, o \textit{speedup} cresce rapidamente. Espera-se que para os testes com mais execuções esse valor seja convergido para um número mais próximo do padrão da curva. Na próxima subseção será apresentado gráficos comparando as performaces para precisão simples e dupla.

Abordando a multiplicação executada 100 e 1000 vezes, nos gráfico das Figuras \ref{fig:gpu_duplo_2} e \ref{fig:gpu_duplo_3}, é possível observar que a quantidade de GFLOPS a partir da ordem $320 \times 320$ se estabiliza em torno de 18. A diferença de tempo de execução, ilustrada nos gráficos das Figuras \ref{fig:gpu_duplo_2b} e \ref{fig:gpu_duplo_3b}, continuam bem alta, como ocorre na precisão simples. Todavia, o tempo de execução da GPU aumenta um pouco quando comparada com o cálculo com \textit{floats} (essa diferença será melhor observada na próxima subseção). Já nos gráficos de \textit{speedup} das \ref{fig:gpu_duplo_2c} e \ref{fig:gpu_duplo_3c}, é possível observar que para a menor matriz é obtido o menor valor, assim como anteriormente, todavia, para ordem $128 \times 128$ o valor se estabiliza em torno de 30 e o \textit{speedup} máximo fica em torno 43, valores abaixo da precisão simples, como já era esperado.

\subsection{Comparação entre precisão simples e dupla}
\begin{figure*}
\centering    
\subfigure[Desempenho em GFLOPS]{
\label{fig:comp_a}
\includegraphics[width=0.6\columnwidth]{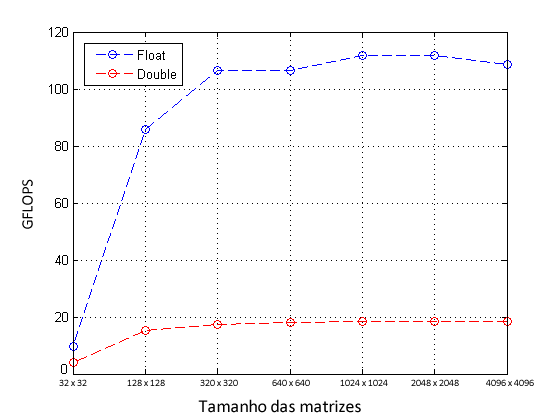}
}
\quad
\subfigure[Desempenho em tempo de execução]{
\label{fig:comp_b}
\includegraphics[width=0.6\columnwidth]{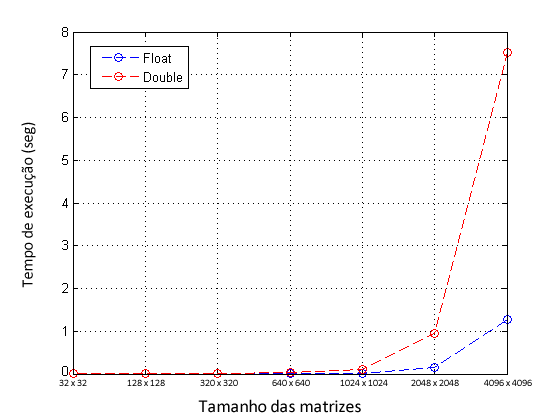}
}
\quad
\subfigure[\textit{Speedup} da CPU + OpenMP em relação a CPU]{
\label{fig:comp_c}
\includegraphics[width=0.6\columnwidth]{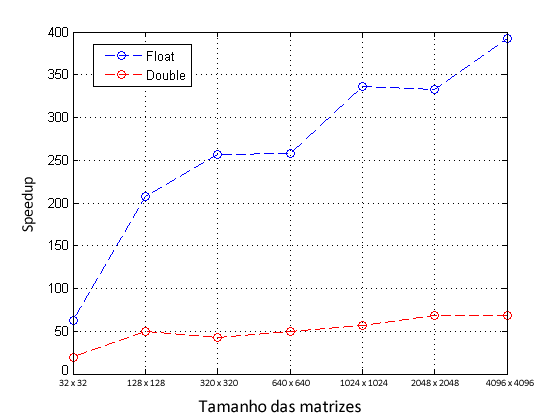}
}

\caption{Comparação do desempenho da GPU para precisão simples e dupla}
\label{fig:comp}
\end{figure*}

Com objetivo de realizar uma breve comparação entre os experimentos com precisão simples e dupla na GPU, são ilustrados na Figura \ref{fig:comp} os gráficos de comparação de GFLOPS, tempo de execução e \textit{speedup} em relação a CPU + OpenMP. Foi utilizado a configuração de 100 execuções. A partir do gráfico na Figura \ref{fig:comp_a}, é possível observar que a partir da multiplicação de matrizes de ordem $128 \times 128$, o processamento da precisão simples cresce substancialmente quando comparada com a dupla. A partir deste ponto, ambas se tornam estáveis. No tempo de execução, ilustrado na Figura \ref{fig:comp_b}, a partir da ordem $1024 \times 1024$ a diferença de tempo já perceptível. Por fim, na comparação de \textit{speedup} em relação a CPU, ilustrada na Figura \ref{fig:comp_c}, é possível perceber que na medida que a ordem das matrizes aumentam, o \textit{speedup} da precisão simples se distancia da precisão dupla.

%% file: src/conclusion.tex
Neste trabalho foi realizado uma comparação entre multiplicação de matrizes utilizando computação paralela e sequencial. A paralelização dos códigos foi alcançada utilizando CUDA, plataforma de codificação em GPU, e OpenMP, \textit{framework} de paralelização em CPU. Para realizar a comparação das abordagens, foram elaborados diversos experimentos utilizando diferentes tamanhos de matrizes. Resumidamente, os resultados dos experimentos mostraram a importância da paralelização ao multiplicar matrizes, principalmente quando o número de elementos das matrizes utilizadas ultrapassa a ordem de 1 milhão. Para casos como este, a GPU se mostrou centenas de vezes mais rápida do que a CPU e a CPU + OpenMP. Também foram analisados a multiplicação com precisão simples e dupla. Neste caso, a paralelização via GPU para precisão simples se mostra mais eficiente do que a dupla pelo fato de sua arquitetura ser privilegiada com número de processadores em relação a dupla. Com isso, vale a pena optar pela precisão simples se a dupla não for um limitante da aplicação. Por fim, ficou claro que para aplicações que utilizam matrizes de grande porte, paralelizar os processos utilizando GPU é uma forma extremamente eficiente de aumentar a performance final.